\documentclass[preprint,aps,floatfix,showpacs]{revtex4}
\usepackage{graphicx}

\newcommand{\bea}{\begin{eqnarray}}
\newcommand{\eea}{\end{eqnarray}}
\newcommand\beq{\begin{equation}}
\newcommand\eeq{\end{equation}}
\newcommand\beqa{\begin{eqnarray}}
\newcommand\eeqa{\end{eqnarray}}

\def\gtap{ \rlap{$>$}\lower5pt\hbox{$\sim$}}

\def\nc{N_c}

\newcommand{\ord}[1]{\ensuremath{\mathcal{O}\left(#1\right)}}

\def\OMIT#1{}

\usepackage{amssymb}

\begin{document}
\preprint{JLAB-THY-05-318}
\vspace{0.5cm}
\title{\phantom{x}
\vspace{0.5cm}  Decays of Non-Strange Positive Parity   Excited  Baryons in
the ${\rm 1/N_C}$ Expansion}

\author{
J. L. Goity $^{a,b}$ \thanks{e-mail: goity@jlab.org} and
N. N. Scoccola $^{c,d,e}$ \thanks{e-mail: scoccola@tandar.cnea.gov.ar}}

\affiliation{
$^a$ Department of Physics, Hampton University, Hampton, VA 23668, USA. \\
$^b$ Thomas Jefferson National Accelerator Facility, Newport News, VA 23606, USA. \\
$^c$ Physics Depart., Comisi\'on Nacional de Energ\'{\i}a At\'omica,
     (1429) Buenos Aires, Argentina.\\
$^d$ CONICET, Rivadavia 1917, (1033) Buenos Aires, Argentina.\\
$^e$ Universidad Favaloro, Sol{\'\i}s 453, (1078) Buenos Aires, Argentina.}
\date{\today}
\begin{abstract}
The decays of non-strange positive parity  excited  baryons via emission
of a  pseudo-scalar meson are studied in the framework of the $1/\nc$ expansion
to order $1/\nc$. In particular, the  pionic  decays of the $\ell=0$
Roper baryons and of the $\ell=2$ baryons in the mass interval 1680-1950 MeV  are analyzed
using  the available partial decay widths.  Decay widths by emission of an $\eta$ meson are
shown to be suppressed by a factor $1/N_c^2$ with respect to the pionic ones.

\end{abstract}

\maketitle

\section{Introduction}
The $1/\nc$ expansion has been  applied  successfully to excited baryons as several studies of
the baryon spectrum  \cite{JLG1,PirjolYan,CCGL,GSS1,CC1,PirjolSchat,Matagne} and decays  \cite{CaroneGeorgi,PirjolYan,CC1,GSS2} have shown.
The  analysis of  baryon masses has shown that there is an approximate $O(3)\times SU(2 N_f)$ symmetry that serves to frame
the $1/\nc$ approach to excited baryons. Although this is not a symmetry that holds strictly in the large $\nc$
limit, its zeroth order in $1/\nc$ breaking turns out to be small. This facilitates the analysis of excited
baryons in general, as one can work  with
a basis of states provided by the multiplets of $O(3)\times SU(2 N_f)$.

The strong decays  represent  one of the  most important problems where the $1/\nc$ expansion can be applied.
Unfortunately the  experimental information on partial widths is neither precise nor complete, and for these reasons the
conclusions that can be drawn from the $1/\nc$ analyses are somewhat limited. The  mesonic decays that have been
so far addressed are
those that proceed via the emission of a single pseudo-scalar meson. The first complete analysis at  $\ord{1/\nc}$ was carried out
for the decays of the  non-strange negative parity baryons \cite{GSS2}.  From  that analysis, the main conclusion  drawn is that the
dominant decay mechanism is provided by the 1-body operator where the pion or $\eta$ meson couples to the axial current as in
the chiral quark model. The deviations from that simple picture are due to a number of operators whose effective couplings are
determined rather poorly because of the mentioned poor quality of the partial width inputs.

The framework used in the present work is similar to the one already developed for the negative-parity baryon decays \cite{GSS2},
which is based on an operator analysis.  A different approach has been recently discussed where the $1/\nc$ expansion is applied
at the level of scattering amplitudes,   and  the baryonic resonances determined from the analytic structure of these
amplitudes \cite{CohenLebed1}.
In the present work the analysis is extended to the known non-strange positive parity  excited  states that fit into the
symmetric representation of $SU(4)$.
As pointed out later, the decays via emission of an $\eta$ meson are sub-leading in $1/\nc$ giving
partial widths  $\ord{1/\nc^2}$.  Consistent with this,  there is virtually  no experimental input available for
a meaningful analysis of the $\eta$ channels.  Thus,  only the decays via emission of a single pion are
 explicitly  considered. The general  motivation for  the present  analysis is to  compare the results with those obtained
 in the negative parity baryons  in order to find out  common features that may underlay the dynamics of the decays in general.

This paper is organized as follows: section II contains the framework for the operator analysis
of the decays, section III presents the numerical results and their analysis,
and finally the conclusions are given in section IV.

\section{ Operator Analysis for Decays  }

In the present application of the $1/N_c$ expansion to
excited baryons we assume that they can be classified in multiplets of the $O(3)\times SU(2N_f)$ group. $O(3)$
corresponds to spatial rotations and $SU(2N_f)$ is the spin-flavor group, where  in this work the number of flavors  is $N_f=2$.
The ground state baryons, namely the $N$ and $\Delta$ states,
belong to the $({\bf 1},{\bf 20_S})$ representation, where the $\bf 20_S$ is the  totally symmetric
representation of $SU(4)$. The positive parity baryons considered here belong instead  to the
$({\bf 2 \ell + 1 },{\bf 20_{S}})$ representation where $\ell$=0,\ 2,\ 4. For general $N_c$,  the excited baryon
states are obtained by coupling the orbital angular momentum state carrying  angular momentum $\ell$ with   the spin-flavor
symmetric states according to:
\begin{equation}
\mid J , J_3 ; I , I_3 ; S \rangle_{exc} =
\sum_m \langle   \ell, m ; S , J_3 - m \mid J, J_3\rangle  \mid \ell , m \rangle
\mid S, J_3 - m ;I=S , I_3 \rangle_{\bf S}.
\end{equation}
The states for  $N_c=3$ are displayed in Table I along with their quantum numbers, masses, decay widths and branching ratios.

The  $\ell_P$ partial wave decay width into a GS baryons  and  a pseudo-scalar meson with isospin $I_P$  is given by
\begin{equation}
\Gamma^{[\ell_P,I_P]}
= \frac{k_P}{8 \pi^2} \frac{M_{B^*}}{M_B} \ \frac{| {\it B} (\ell_P,I_P,S,I,J^*,I^*,S^*) |^2}
{\sqrt{(2 J^* + 1)(2 I^*+1)}},
\label{width}
\end{equation}
where ${\it B} (\ell_P,I_P,S,I,J^*,I^*,S^*)$ are the reduced matrix elements of the baryonic
operator. Such operator admits an expansion in $1/N_c$ that has the general form \cite{GSS2}:
\begin{equation}
B^{[\ell_P, I_P]}_{[\mu, \alpha]} =
\left(\frac{k_P}{\Lambda}\right)^{\ell_P}\sum_q \, C_q^{[\ell_P, I_P]}(k_P)
\left( B^{[\ell_P, I_P]}_{[\mu, \alpha]} \right)_q,
\label{exp}
\end{equation}
where
\begin{equation}
\left( B^{[\ell_P, I_P]}_{[\mu, \alpha]} \right)_q = \sum_m
\langle \ell, m ; j, j_z  \mid \ell_P, \mu \rangle \
\xi^\ell_m \ \left( {\cal G}^{[j, I_P]}_{[j_z,\alpha]} \right)_q.
\end{equation}
The factor  $\left(\frac{k_P}{\Lambda}\right)^{\ell_P}$ is included to take
into account the chief meson momentum dependence of the partial wave.  For definiteness, in the following the scale $\Lambda$ is taken to be equal to 200 MeV.  The operator  $\xi^\ell_m$  drives the transition from the $(2\ell+1)$-plet to the singlet $O(3)$ state,
and the spin-flavor operators  $\left( {\cal G}^{[j, I_P]}_{[j_z,\alpha]} \right)_q$  give  transitions within the  symmetric $SU(4)$ representation in which both the excited and GS baryons reside. The label  $j$ denotes  the spin of the spin-flavor operator while its  isospin obviously coincides with the isospin of the emitted meson.
The dynamics of the decays is encoded in the effective dimensionless coefficients
$C_q^{[\ell_P, I_P]}(k_P)$. The reduced matrix elements of the operators $\left( B^{[\ell_P, I_P]}_{[\mu, \alpha]} \right)_q$ appearing
in Eq.(\ref{exp})
can be easily calculated in terms of the reduced matrix elements of the spin-flavor operators.
Note that in the present case, where $\ell_P$ can be 1, 3 or 5 only
(i.e. P, F or H waves)
\footnote{For $\ell = 4$ there is, in addition, a pion J-wave decay which we are
not going to consider here.}, the spin-flavor operators can carry spin $j=1,\cdots,5$.

The terms on the right hand side of Eq.(\ref{exp}) are ordered in powers of $1/N_c$.
The order in $1/N_c$ is determined by the spin-flavor operator, where for an $n$-body operator the power  is  given by \cite{DJM}
\begin{equation}
\nu=n-1-\kappa,
\end{equation}
where the power $n-1$ results from the number of gluon exchanges needed to generate an $n$-body operator at the quark level.
The coherence factor  $\kappa$ is equal to zero for incoherent operators and can be equal to one or even larger (up to $n$)
for coherent operators.

The effective operators are defined such that  all  coefficients
$C_q^{[\ell_P, I_P]}(k_P)$ in Eq.(\ref{exp}) are of zeroth order in $N_c$.
The leading order of the
decay amplitude is in fact $N_c^0$  or higher, depending on the channel considered. 
 \cite{GSS2,GoityMixDecay}.
At this point it is opportune to discuss  the momentum dependence of
the coefficients. The  spin-flavor breaking  in the masses,  in both excited
and GS baryons,  give rise to different values of the momenta $k_P$.
In this work, we adopt a scheme where the only
momentum dependence assigned to the coefficients is the
explicitly shown factor    $\left(k_P/\Lambda\right)^{\ell_P}$ that
takes into account the chief momentum dependence of the
partial wave,  while the rest of the dependence is absorbed  into  the coefficients
of the sub-leading operators.


The construction of transition operators follows similar steps
as in the decays of negative parity baryons \cite{GSS2}. In the present case, however, there is an important
simplification because the  transitions only involve states  in symmetric spin-flavor
representations.  The simplification is that there is no need to distinguish between excited
and core spin-flavor operators, and thus  an arbitrary transition operator can be
constructed simply in terms of appropriate   products of the generators of the spin-flavor group
\begin{eqnarray}
\Lambda_\gamma &=& S^{[1,0]},\ T^{[0,1]}, \ G^{[1,1]}, \label{transbasic}
\end{eqnarray}
where   the notation $O^{[j,t]}$ is used to indicate the spin $j$ and the isospin $t$ of the operators.

A composite spin-flavor operator of spin $j$ must be at least $j$-body.
 For arbitrary $N_c$, only $n$-body operators with $n \leq N_c$ are allowed.
However, since in the present analysis one takes in the end $\nc=3$, only composite operators up to at most
3-body are needed.
The composite operators that provide the basis for the decay amplitudes  are
constructed by coupling the composite spin-flavor operators to the
$\xi^{[\ell,0]}$ operator that acts on the $O(3)$ degrees of freedom of the excited quark.
The composite spin-flavor operators  up to 3-body ones are of the form:
\begin{equation}
\frac{1}{N_c} \Lambda_\gamma
\qquad ; \qquad
\frac{1}{N_c^2} \ \{ \Lambda_{\gamma_1},\ \Lambda_{\gamma_2}\}
\qquad ; \qquad
\frac{1}{N_c^3} \ \{\Lambda_{\gamma_1},\{ \Lambda_{\gamma_2}, \Lambda_{\gamma_3}\}\}.
\label{123B}
\end{equation}
where the factors $1/N_c^n$ take into account   the usual gluon
exchange rules mentioned earlier.
For decays in the pion channels these products transform as $[j,1]$, where $j=1,2,3$
(for general $\nc$ our analysis would have included also $j=4,5$ that involve $n$-body operators with $n > 3$).
For decays in the $\eta$ channel they transform instead as  $[j,0]$.

  From
the transformation properties of the generators given in Eq.(\ref{transbasic}) it is straightforward to
construct all possible products of the forms given in Eqs.(\ref{123B})
with $[j,1]$, where $j=1,2,3$. The rather long list of operators obtained
in this way can be reduced by using the reduction  rules  that  apply to
matrix elements of products of generators  in the symmetric representation \cite{DJM}. The reduced list that results from keeping contributions  that are  at most order
$1/N_c$ is
\begin{equation}
\frac{1}{N_c} G
\qquad ; \qquad
\frac{1}{N_c^2} \ \left( [ S, G ] \right)^{[1,1]}
\qquad ; \qquad
\frac{1}{N_c^2} \ \left( [ S, G ] \right)^{[2,1]}
\qquad ; \qquad
\frac{1}{N_c^3} \ \left( G ( G G )^{[2,2]} \right)^{[j,1]},
\end{equation}
where  $( [ S, G ] )^{[1,1]} $ denotes $ (S G)^{[1,1]} - (G S)^{[1,1]}
= \langle 1 \alpha 1 \beta | 1 \delta \rangle ( S^{\alpha}  G^{\beta a} - G^{\alpha a} S^{\beta} )$, etc..
Since $G$ is a coherent operator, one can see that the 1-body operator is $\ord{\nc^0}$, while the rest are $\ord{1/\nc}$.

It is rather straightforward to build the composite spin-flavor operators that are needed to describe decays in the $\eta$ channel.
For instance, the 1-body operator is $S/\nc$, which is $\ord{1/\nc}$. One can immediately establish that in fact all the operators
are suppressed, and therefore the $\eta$ channel
is in amplitude suppressed by a factor $1/\nc$ with respect to the pion channel.

Further reductions in the number of operators result from the fact that
not all of the  operators are linearly independent at  order $1/N_c$. The determination of  the final set of independent operators
for each particular decay channel is more laborious. This is achieved by coupling the  spin-flavor operators with  the
$\xi^{[\ell,0]}$ orbital transition operator to the corresponding
total spin and isospin and by explicitly calculating  all the  matrix elements.
In this way,  operators that are linearly dependent at the corresponding order in the $1/N_c$ expansion can be eliminated.
The   basis of independent operators
$\left( O^{[\ell_P, I_P]}_{[m_P,I_{P_3}]}\right)_n$ is displayed in Table~II, where for the sake of
simplicity the spin and isospin projections have been omitted.  It is interesting to observe that in the end
only 1- and 2-body operators are left.  Note  also  that in the case of the $\ell=0$ baryons
$O_3^{[\ell_P,1]}$ is absent.
The  corresponding reduced matrix elements are given in Tables III through V.
In the bottom rows of these tables normalization coefficients
$\alpha^{[\ell,I_P]}_n$ are given. These coefficients are
used to define normalized basis operators
such that, for $N_c=3$, their largest
reduced matrix element is equal to one for order $N_c^0$ operators
and equal to 1/3 for order $1/N_c$ operators. Thus,
\begin{equation}
\left( B^{[\ell_P, I_P]}_{[m_P,I_{P_3}]}\right)_n =
\alpha^{[\ell,I_P]}_n  \left( O^{[\ell_P, I_P]}_{[m_P,I_{P_3}]}\right)_n
\end{equation}
furnishes the list of basis operators normalized according to the $1/N_c$ power counting.

\section{Results}

 In the context of the $1/N_c$ expansion,  the  decays of the Roper multiplet were already studied  in Ref. \cite{CC1}, where
the whole $SU(6)$ 56-plet was analyzed using only the 1-body  operator $O_1^{[0,1]}$ and the breaking of spin and flavor
symmetries was included by means of a  profile function conveniently chosen and adjusted by a best fit to the partial widths.
Here we analyze the P-wave decays (for the F-wave decay of the $\Delta(1600)$ there is too little to be said).  The analysis
includes  the $1/N_c$ corrections via the only operator  $\ell_P=0$  that appears at that order, namely the 2-body operator
$O_2^{[0,1]}$, and do not include a profile function. The results of the fits are shown in  Table VI, where it is evident
that  at LO  the
fit turns out to be  rather poor.  At NLO on the other hand the fit is excellent. The sub-leading operator is
essential  for inverting the ordering of the partial widths of the $\Delta(1600)$   obtained  at LO.
The  NLO  corrections are also essential for improving the ratio of the two $N(1440)$ partial widths,
which is reproduced rather poorly at LO.
 The coefficient of the LO operator remains stable as one includes the NLO corrections, and the
coefficient of the NLO operator is of natural size when compared with the coefficient of the LO
operator.


In the  case  of the $\ell=2$ states  one
has two  nucleon and four  $\Delta$ states, and the relevant decays can be P- and F-wave. In both channels there are three
operators when the $1/\nc$ corrections are included. There are only four P-wave partial widths  listed by the
Particle Data Group \cite{PDG},  which  is sufficient for the analysis to $\ord{1/\nc}$.  As Table VII shows, in this case one finds  that
an excellent fit is already obtained at LO.
This  means that the NLO operators are of
marginal significance. In fact, in the NLO fit the  coefficients are compatible with zero.
For the F-wave decays there are five known widths,  two of them ($N(1680)\rightarrow \pi \ N$ and
$\Delta(1950)\rightarrow \pi \ N$) are known with an accuracy of about $10 \%$,  which  is  the magnitude  of  the  $\ord{1/\nc^2}$ corrections.  
Thus, in order to perform a consistent LO fit those
empirical errors of 10\% are replaced by 30\% errors, which is the magnitude of the $\ord{1/N_c }$ corrections.  The results in Table
VIII show that a reasonably good LO fit is obtained in this way. On the other hand,  in the  NLO fits
the errors used are the ones  given by  the Particle Data Group \cite{PDG}. Several of the errors  remain around  30\%, which is poor accuracy for inputs to a NLO fit.  Two possible NLO fits are given in Table VIII. In the \#1~NLO fit all five experimental inputs are included resulting in a very large value
of $\chi^2$. It is found that the narrow partial width assigned to $N(1680)\to \pi\Delta$ is the
chief
contribution to the $\chi^2$ in this fit. It should be noticed that the PDG only quotes an upper bound
for the corresponding branching ratio. If one disregards that input (\#2~NLO fit),  an excellent fit is obtained
where one of the  NLO coefficients is of  natural size and the other one associated with he operator $O_3^{[2,1]}$
is suppressed.  Actually,  an equally  good fit can be obtained by disregarding the latter operator. Interestingly
enough, the predicted $N(1680)\to \pi\Delta$ width is predicted to be  several times smaller
than the rest of the F-wave pion decay widths, which  qualitatively agrees with observation despite violating
the experimental bound slightly.

Finally, for the $\ell=4$ baryons the available experimental information is unfortunately too incomplete to warrant
any reasonable analysis. In fact, there is no data about the F-wave decays, and there is  information about only two out of
nine possible H-wave decays. In any case, as soon as more information will become available Eqs.(\ref{width}, \ref{exp})
together with the expressions in Table V could be readily used  to perform the corresponding analysis.

\section{Conclusions}
The decays via  pseudo-scalar meson  emission of the  low lying  positive parity   excited baryons
have been studied to $\ord{1/\nc}$.  For pionic decays of the $\ell = 0$ Roper baryons  it is found that
the $1/\nc$ corrections are essential for obtaining a good fit.  The  $1/\nc$ corrections are  clearly  required
by the observed ordering of the two $\Delta(1600)$ partial widths, namely
$\Gamma(\Delta(1600)\to \pi N)<\Gamma(\Delta(1600)\to \pi \Delta)$, which
at LO cannot be accounted for. It is important to notice that these are the only decays
where the LO 1-body operator $O_1$ (see table II), that is the dominant one in a chiral quark model picture,
does not give a qualitatively correct picture of the decays.  The meaning of this observation for the nature
of the Roper multiplet is an interesting issue to understand. The  pionic  decays of the $\ell=2$ multiplet are
qualitatively well described at LO, and the improvements at NLO are accomplished with essentially one extra operator
whose coefficient is of natural magnitude. As in the case of the negative parity baryons,  the extension of the analyses
to include the strange excited baryons  seems to be the natural step to gain further insight on decays. The general point that one
should emphasize is that so far the main limitation in the analyses is due to the poor quality of the input partial widths.
To improve this it is necessary to have extensive experimental progress, which is in part being fulfilled by the present $N^*$
program at Jefferson Lab \cite{Burkert}.

One interesting observation that points to the consistency of the framework based on the approximate $O(3)\times SU(4)$
symmetry is that the predicted suppression of the $\eta$ channels in the decays discussed here is clearly displayed by
the observed decays. This represents a strong experimental confirmation that the excited baryons we analyzed do belong
primarily into a symmetric representation of $SU(4)$. In the case of the negative parity baryons the $\eta$ channel
is not suppressed and is indeed very important. This implies that these states belong primarily  into the mixed-symmetric
representation of $SU(4)$. Thus, $\eta$ channels serve as a selector of the spin-flavor structure of excited baryons.

\section*{ACKNOWLEDGEMENTS}

This work was supported by DOE contract DE-AC05-84ER40150 under which SURA operates the Thomas Jefferson
National Accelerator Facility,  by the National Science Foundation (USA) through grants \#~PHY-9733343 and -0300185 (JLG),
and  by CONICET (Argentina) grant \# PIP 02368 and by ANPCyT  (Argentina) grant \# PICT 00-03-08580 (NNS).

\pagebreak

\begin{table}[htdp]
\caption{Positive parity non-strange baryons and their decay
widths and branching ratios from the PDG. Channels not explicitly
indicated are forbidden. Question marks imply that the empirical
value is not known. Channels for which only one decay is allowed
are not considered.}
\begin{center}
\narrowtext
\begin{tabular}{ccccccccc} \hline \hline
\hspace*{.3cm} $B_J^\ell$ \hspace*{.3cm}&
\hspace*{.3cm}State \hspace*{.3cm}&  \hspace*{.3cm}Mass \hspace*{.3cm}
& \hspace*{.3cm}Total width\hspace*{.3cm} & \multicolumn{4}{c}{Branching ratios [\%]} \\
    \cline{5-8}
  &&[MeV]&[MeV]& & \hspace*{.3cm} P-wave\hspace*{.3cm} & \hspace*{.3cm} F-wave \hspace*{.3cm}
  & \hspace*{.3cm} H -wave \hspace*{.3cm}\\ \hline\hline
 $N^0_{1/2}$      & $N(1440)$      &   $1450\pm 20$    & $350 \pm 100 $ &   $\pi N$     & $65\pm 5$      &                 & \\[-2.5mm]
                  &                &                   &                &   $\pi \Delta$& $25\pm 5$      &                 & \\
\hline
$\Delta^0_{3/2}$  & $\Delta(1600)$ &   $1625\pm 75$    & $350 \pm 100$  &   $\pi N$     & $17.5\pm 7.5$  &                 & \\[-2.5mm]
                  &                &                   &                &   $\pi \Delta$& $51.5\pm 24$   &  Not considered & \\
\hline \hline
  $N^2_{3/2}$       & $N(1720)$    &    $1700\pm 50$  & $150\pm 50$    &   $\pi N$      & $15.5\pm 5$    &                 & \\[-2.5mm]
                  &                &                  &                &   $\pi \Delta$ & $ ? $          &        $?$      & \\
\hline
  $N^2_{5/2}$       & $N(1680)$       &  $1683\pm 8$   & $130\pm 10$    &   $\pi N$     &                &    $65\pm 5$    & \\[-2.5mm]
                  &                 &                  &                &   $\pi \Delta$& $10\pm 4 $     &      $1\pm1$    & \\
\hline
  $\Delta^2_{1/2}$& $\Delta(1910)$  &  $1895\pm 25$    & $230\pm 40$    &   $\pi N$     &  $22.5\pm 7.5$ &                 &  \\[-2.5mm]
                  &                 &                  &                &   $\pi \Delta$&  $?$           &                 & \\
\hline
  $\Delta^2_{3/2}$&  $\Delta(1920)$ &  $1935\pm 35$    & $225\pm 75$    &   $\pi N$     &  $12.5\pm 7.5$ &                 &   \\[-2.5mm]
                  &                 &                  &                &   $\pi \Delta$&  $?$           &  $?$            &   \\
\hline
  $\Delta^2_{5/2}$&  $\Delta(1905)$ &  $1895\pm 25$    & $360\pm 80$    &   $\pi N$     &                &   $10\pm 5$     &  \\[-2.5mm]
                  &                 &                  &                &   $\pi \Delta$&  $?$           &   $?$           &  \\
\hline
  $\Delta^2_{7/2}$&  $\Delta(1950)$ &  $1950\pm 10$    & $320\pm 30$    &   $\pi N$     &                &  $37.5\pm 2.5$  & \\[-2.5mm]
                  &                 &                  &                &   $\pi \Delta$&                &  $24\pm 4$      & Not considered \\
\hline \hline
  $N^4_{7/2}$     & $N(?)$          &  $?$             & $?$            &   $\pi N$     &                &   $?$           &       \\[-2.5mm]
                  &                 &                  &                &   $\pi \Delta$&                &                 &  $?$  \\
\hline
  $N^4_{9/2}$     & $N(2220)$       &  $2245\pm 65$    & $450\pm 100$   &   $\pi N$     &                &                 &   $15\pm 5$  \\[-2.5mm]
                  &                 &                  &                &   $\pi \Delta$&                &    $?$          &   $?$        \\
\hline
  $\Delta^4_{5/2}$& $\Delta(?)$  &  $?$              & $?$              &   $\pi N$     &                &    $?$         &                     \\[-2.5mm]
                  &              &                   &                  &   $\pi \Delta$&                &    $?$         &                    \\
\hline
  $\Delta^4_{7/2}$&  $\Delta(2390)^*$ &  $2390\pm ?$   & $?$         &      $\pi N$     &                &                &    $?$       \\[-2.5mm]
                  &                 &                  &             &      $\pi \Delta$&                &    $?$         &    $?$       \\
\hline
$\Delta^4_{9/2}$  &$\Delta(2300)^{**}$&  $2300\pm ?$   & $?$            &   $\pi N$     &                &                &     $?$   \\[-2.5mm]
                  &                 &                  &                &   $\pi \Delta$&                &     $?$        &     $?$   \\
\hline
$\Delta^4_{11/2}$ &  $\Delta(2420)$ &  $2400\pm 100$   & $400\pm 100$   &   $\pi N$     &                &                &   $10\pm 5$  \\[-2.5mm]
                  &                 &                  &                &   $\pi \Delta$&                &                &   $?$   \\
\hline \hline

\end{tabular}
\end{center}
\label{default}
\end{table}

\pagebreak

\begin{table}[ht]
\begin{center}
\caption{Basis operators. Note that in the case $\ell=0$ the operator
$O_3^{[\ell_P,1]}$ is absent.}
\vspace*{1cm}
\begin{tabular}{cccccccccc}\hline\hline
\hspace*{.3cm} 
\hspace*{.3cm}   &
\hspace*{.3cm} Name                \hspace*{.3cm}   &
\hspace*{.3cm} Operator            \hspace*{.3cm}   &
\hspace*{.3cm} Order in $1/N_c$    \hspace*{.3cm}
\\ \hline \hline
 1B   &  $O_1^{[\ell_P,1]}$ &
$\frac{1}{N_c}\left( \xi^\ell \ G \right)^{[\ell_P,1]}$
& 0 \\[2mm] \hline \hline
 2B  &   $O_2^{[\ell_P,1]}$ &
$\frac{1}{N_c^2} \left(\xi^\ell \left( [ S \ , \ G ] \right)^{[1,1]} \right)^{[\ell_P,1]}$
& 1 \\[2mm]
     &   $O_3^{[\ell_P,1]}$ &
$\frac{1}{N_c^2} \left(\xi^\ell \left( \{ S \ , \ G \} \right)^{[2,1]} \right)^{[\ell_P,1]}$
& 1 \\[2mm] \hline \hline
\end{tabular}
\end{center}
\end{table}

\begin{table}[ht]
\begin{center}
\caption{Reduced matrix element of basis operators for the P wave decays of $\ell=0$ excited baryons.}
\vspace*{1cm}
\begin{tabular}{ccccc}
\hline\hline
Pion P waves  &\hspace*{.3cm} $O_1^{[1,1]}$\hspace*{.3cm}  &
               \hspace*{.3cm} $O_2^{[1,1]}$\hspace*{.3cm}  &
               \hspace*{.5cm} Overall factor \hspace*{.5cm} \\
\hline\hline
   $N^0_{1/2}\rightarrow \pi \ N$
         & $1$                      & $0$                   &
 $-\frac{N_c+2}{2 N_c}$ \\
   $N^0_{1/2}\rightarrow \pi \ \Delta$
         & $1$      & $-\frac{3}{\sqrt{2} N_c}$    &
 $-\frac{\sqrt{(N_c+5)(N_c-1)}}{\sqrt{2} N_c}$\\
   $\Delta^0_{3/2}\rightarrow \pi \ \Delta$
         & $1$     & $\frac{3}{\sqrt{2} N_c}$     &
 $-\frac{\sqrt{(N_c+5)(N_c-1)}}{\sqrt{2} N_c}$ \\
   $\Delta^0_{3/2}\rightarrow \pi \ N$
         & $1$       & $0$     &
 $-\frac{N_c+2}{N_c}$ \\ \hline
$\alpha$    & $-\frac35$   & $\frac12$ &     &                 \\
\hline \hline
\end{tabular}
\end{center}
\end{table}

\begin{table}[h]
\begin{center}
\caption{Reduced matrix element of basis operators for the decays of $\ell=2$ excited baryons}
\vspace*{1cm}
\begin{tabular}{cccccc}
\hline\hline
Pion P waves  &\hspace*{.3cm} $O_1^{[1,1]}$\hspace*{.3cm}  &
               \hspace*{.3cm} $O_2^{[1,1]}$\hspace*{.3cm}  &
               \hspace*{.3cm} $O_3^{[1,1]}$\hspace*{.3cm}  &
               \hspace*{.5cm} Overall factor \hspace*{.5cm} \\
\hline\hline
   $N^2_{3/2}\rightarrow \pi \ N$
         & $1$                      & $0$                   & $0$                      &
 $\frac{N_c+2}{2 N_c}$ \\
   $N^2_{3/2}\rightarrow \pi \ \Delta$
         & $\frac{1}{\sqrt2}$      & $-\frac{3}{2 N_c}$    & $\frac{3\sqrt3}{2 N_c}$  &
 $-\frac{\sqrt{(N_c+5)(N_c-1)}}{\sqrt{10} N_c}$\\
   $N^2_{5/2}\rightarrow \pi \ \Delta$
         & $-\frac{1}{2\sqrt2}$     & $\frac{3}{4 N_c}$     & $\frac{1}{4\sqrt3 N_c}$  &
 $3\sqrt{\frac{2}{5}} \frac{\sqrt{(N_c+5)(N_c-1)}}{N_c}$ \\
   $\Delta^2_{1/2}\rightarrow \pi \ N$
         & $\frac{1}{\sqrt2}$       & $\frac{3}{2 N_c}$     & $\frac{\sqrt3}{2 N_c}$   &
 $-\frac{\sqrt{(N_c+5)(N_c-1)}}{\sqrt2 N_c}$\\
   $\Delta^2_{1/2}\rightarrow \pi \ \Delta$
         & $-\frac{1}{\sqrt3}$      & $0$                   & $-\frac{2 \sqrt2}{N_c}$    &
 $-\sqrt{\frac{3}{5}} \frac{N_c+2}{2N_c}$\\
   $\Delta^2_{3/2}\rightarrow \pi \ N$
         & $-\frac{1}{\sqrt2}$      & $-\frac{3}{2 N_c}$    & $\frac{\sqrt3}{2 N_c}$   &
 $\frac{\sqrt{(N_c+5)(N_c-1)}}{\sqrt2 N_c}$\\
   $\Delta^2_{3/2}\rightarrow \pi \ \Delta$
         & $\frac{1}{\sqrt2}$       & $0$                   & $\frac{\sqrt3}{ N_c}$    &
 $\frac{4}{5} \frac{N_c+2}{N_c}$\\
   $\Delta^2_{5/2}\rightarrow \pi \ \Delta$
         & $-3$                     & $0$                   & $\frac{2 \sqrt6}{N_c}$   &
 $-\frac{\sqrt7}{10} \frac{N_c+2}{N_c}$\\
\hline
$\alpha$    & $\frac{2}{\sqrt7}$   & $\frac{\sqrt{10}}{6}$ &
$\frac{\sqrt3}{4}$  &                 \\ \hline
            &                                 &                       &
                               &                 \\
\hline
\hline
Pion F waves  & $O_1^{[3,1]}$  &  $O_2^{[3,1]}$ & $O_3^{[3,1]}$  & Overall factor \\
\hline\hline
   $N^2_{3/2}\rightarrow \pi \ \Delta$
         & $1$            & $-\frac{3}{\sqrt2 N_c}$   & $\frac{1}{N_c}$        &
 $-\sqrt{\frac{7}{10}} \frac{\sqrt{(N_c+5)(N_c-1)}}{N_c}$ \\
   $N^2_{3/2}\rightarrow \pi \ N$
         & $1$            & $0$                        & $0$                      &
 $-\sqrt{\frac{7}{3}} \frac{N_c+2}{2 N_c}$              \\
   $N^2_{5/2}\rightarrow \pi \ \Delta$
         & $1$            & $-\frac{3}{\sqrt2 N_c}$   & $-\frac{3}{2 N_c}$       &
 $-\sqrt{\frac{7}{15}} \frac{\sqrt{(N_c+5)(N_c-1)}}{N_c}$ \\
   $\Delta^2_{3/2}\rightarrow \pi \ \Delta$
         & $1$            & $0$                        & $\frac{4}{N_c}$          &
 $-\frac{\sqrt7}{5} \frac{N_c+2}{N_c}$                  \\
   $\Delta^2_{5/2}\rightarrow \pi \ N$
         & $1$            & $\frac{3}{\sqrt2 N_c}$    & $\frac{3}{N_c}$        &
 $-\frac{1}{\sqrt6} \frac{\sqrt{(N_c+5)(N_c-1)}}{N_c}$  \\
   $\Delta^2_{5/2}\rightarrow \pi \ \Delta$
         & $1$            & $0$                        & $\frac{3}{2 N_c}$        &
 $-\frac{8}{5\sqrt3}\frac{N_c+2}{N_c} $                 \\
   $\Delta^2_{7/2}\rightarrow \pi \ N$
         & $1$            & $\frac{3}{\sqrt2 N_c}$    & $-\frac{1}{2 N_c}$       &
 $-\frac{\sqrt{(N_c+5)(N_c-1)}}{N_c}$                   \\
   $\Delta^2_{7/2}\rightarrow \pi \ \Delta$
         & $1$            & $0$                        & $-\frac{2}{N_c}$         &
 $-\sqrt{\frac{6}{5}} \frac{N_c+2}{N_c}$                \\
\hline
$\alpha$ & $-\sqrt{\frac{3}{10}}$   & $-\frac{1}{2\sqrt2}$ &  $\frac{1}{2}\sqrt{\frac{3}{10}}$  &
\\ \hline\hline
\end{tabular}
\end{center}
\end{table}

\begin{table}[h]
\begin{center}
\caption{Reduced matrix element of basis operators for the decays of $\ell=4$ excited baryons}
\vspace*{1cm}
\begin{tabular}{cccccc}
\hline\hline
Pion F waves  &\hspace*{.3cm} $O_1^{[3,1]}$\hspace*{.3cm}  &
               \hspace*{.3cm} $O_2^{[3,1]}$\hspace*{.3cm}  &
               \hspace*{.3cm} $O_3^{[3,1]}$\hspace*{.3cm}  &
               \hspace*{.5cm} Overall factor \hspace*{.5cm} \\
\hline\hline
   $N^4_{7/2}\rightarrow \pi \ N$
         & $1$                      & $0$                   & $0$                      &
 $\sqrt{\frac73}\frac{N_c+2}{2 N_c}$ \\
   $N^4_{7/2}\rightarrow \pi \ \Delta$
         & $1$      & $-\frac{3}{\sqrt2 N_c}$    & $\frac{\sqrt{15}}{\sqrt2 N_c}$  &
 $-\frac{\sqrt7\sqrt{(N_c+5)(N_c-1)}}{6 N_c}$\\
   $N^4_{9/2}\rightarrow \pi \ \Delta$
         & $1$     & $-\frac{3}{\sqrt2 N_c}$     & $-\frac{\sqrt3}{\sqrt{10} N_c}$  &
 $-\frac{\sqrt{35} \sqrt{(N_c+5)(N_c-1)}}{6 N_c}$ \\
   $\Delta^4_{5/2}\rightarrow \pi \ N$
         & $1$       & $\frac{3}{\sqrt2 N_c}$     & $\frac{\sqrt5}{\sqrt6 N_c}$   &
 $-\frac{\sqrt3\sqrt{(N_c+5)(N_c-1)}}{2 N_c}$\\
   $\Delta^4_{5/2}\rightarrow \pi \ \Delta$
         & $1$      & $0$                   & $2\frac{\sqrt{10}}{\sqrt3 N_c}$    &
 $\sqrt{\frac{3}{2}} \frac{N_c+2}{2N_c}$\\
   $\Delta^4_{7/2}\rightarrow \pi \ N$
         & $1$      & $\frac{3}{\sqrt2 N_c}$    & $-3\frac{\sqrt3}{\sqrt{10} N_c}$   &
 $-\frac{\sqrt5\sqrt{(N_c+5)(N_c-1)}}{\sqrt{12} N_c}$\\
   $\Delta^4_{7/2}\rightarrow \pi \ \Delta$
         & $1$       & $0$                    & $\frac{\sqrt6}{\sqrt5 N_c}$    &
 $\frac{2\sqrt2}{3} \frac{N_c+2}{N_c}$\\
   $\Delta^4_{9/2}\rightarrow \pi \ \Delta$
         & $1$                     & $0$      & $-2\frac{\sqrt6}{\sqrt5 N_c}$   &
 $\sqrt{\frac{77}{18}} \frac{N_c+2}{2 N_c}$\\
\hline
$\alpha$    & $\frac{18\sqrt2}{5\sqrt{77}}$   & $\frac{3}{\sqrt{70}}$ &
$-\frac{3\sqrt3}{\sqrt{385}}$  &                 \\ \hline
            &                                 &                       &
                               &                 \\
\hline
\hline
Pion H waves  & $O_1^{[5,1]}$  &  $O_2^{[5,1]}$ & $O_3^{[5,1]}$  & Overall factor \\
\hline\hline
   $N^4_{7/2}\rightarrow \pi \ \Delta$
         & $1$            & $-\frac{3}{\sqrt2 N_c}$   & $\frac{\sqrt3}{2 N_c}$        &
 $-\frac{\sqrt{11}}{3} \frac{\sqrt{(N_c+5)(N_c-1)}}{N_c}$ \\
   $N^4_{9/2}\rightarrow \pi \ N$
         & $1$            & $0$                        & $0$                      &
 $-\sqrt{\frac{11}{3}} \frac{N_c+2}{2 N_c}$              \\
   $N^4_{9/2}\rightarrow \pi \ \Delta$
         & $1$            & $-\frac{3}{\sqrt2 N_c}$   & $-\frac{\sqrt3}{N_c}$       &
 $-\sqrt{\frac{11}{2}} \frac{\sqrt{(N_c+5)(N_c-1)}}{3 N_c}$ \\
   $\Delta^4_{7/2}\rightarrow \pi \ \Delta$
         & $1$            & $0$                        & $\frac{2\sqrt3}{N_c}$          &
 $-\frac{\sqrt{154}}{15} \frac{N_c+2}{N_c}$                  \\
   $\Delta^4_{9/2}\rightarrow \pi \ N$
         & $1$            & $\frac{3}{\sqrt2 N_c}$    & $\frac{3\sqrt3}{2 N_c}$        &
 $-\frac{1}{\sqrt3} \frac{\sqrt{(N_c+5)(N_c-1)}}{N_c}$  \\
   $\Delta^4_{9/2}\rightarrow \pi \ \Delta$
         & $1$            & $0$                        & $\frac{\sqrt3}{2 N_c}$        &
 $-\frac{8}{\sqrt5}\frac{N_c+2}{3 N_c} $                 \\
   $\Delta^4_{11/2}\rightarrow \pi \ N$
         & $1$            & $\frac{3}{\sqrt2 N_c}$    & $-\frac{1}{\sqrt3 N_c}$       &
 $-\sqrt{\frac32}\frac{\sqrt{(N_c+5)(N_c-1)}}{N_c}$                   \\
   $\Delta^4_{11/2}\rightarrow \pi \ \Delta$
         & $1$            & $0$                        & $-\frac{4}{\sqrt3 N_c}$         &
 $-\frac{\sqrt{39}}{5} \frac{N_c+2}{N_c}$                \\
\hline
$\alpha$ & $-\sqrt{\frac{3}{13}}$   & $-\frac{1}{2\sqrt3}$ &  $\frac{3}{4\sqrt{13}}$  &
\\ \hline\hline
\end{tabular}
\end{center}
\end{table}

\begin{table}[h]
\begin{center}
\caption{Fit parameters and partial widths corresponding to the pion P wave decays of
$\ell=0$ excited baryons}
\vspace*{1cm}
\begin{tabular}{ccccc}
\hline\hline
Pion P waves  & \hspace*{.5cm} Emp. Width \hspace*{.5cm}&\hspace*{.5cm} LO  \hspace*{.7cm}& \hspace*{.7cm} NLO \hspace*{.5cm}  \\
              &     MeV     &   MeV   &   MeV     \\
\hline
$\chi^2_{dof}$&             & 4.05    &    0.1      \\
dof           &             &   3     &    2      \\
\hline
$C_1^{[1,1]}$ &             & $18.7\pm 2.4$    &  $ 17.0\pm 1.6$   \\
$C_2^{[1,1]}$ &             &  -               &  $ 24.4\pm 6.3$   \\
\hline
   $N(1440)\rightarrow \pi \ N$
    & $227.5 \pm 67.3$   &   106      &    245     \\
   $N(1440)\rightarrow \pi \ \Delta$
   &  $87.5\pm 30.5$     &    16.0    &    83.6     \\
   $\Delta(1600)\rightarrow \pi \ N$
   & $61.25\pm 31.6$     &   106      &    59.4    \\
   $\Delta(1600)\rightarrow \pi \ \Delta$
   & $180\pm 99$         &    63.5    &   146      \\
\hline
\hline
\end{tabular}
\end{center}
\end{table}

\begin{table}[h]
\begin{center}
\caption{Fit parameters and partial widths corresponding to the pion P wave decays of
$\ell=2$ excited baryons}
\vspace*{1cm}
\begin{tabular}{ccccc}
\hline\hline
Pion P waves  & \hspace*{.5cm} Emp. Width \hspace*{.5cm}& \hspace*{.5cm}  LO  \hspace*{.8cm} & \hspace*{.8cm} NLO  \hspace*{.5cm} \\
              &     MeV     &   MeV   &   MeV    \\
\hline
$\chi^2_{dof}$&             & 0.15    &    0.44  \\
dof           &             &   3     &    1     \\
\hline
$C_1^{[1,1]}$ &             & $6.83\pm 0.77$  & $4.09\pm 0.47$   \\
$C_2^{[1,1]}$ &             &  -              & $0.11\pm 2.41$   \\
$C_3^{[1,1]}$ &             &  -              & $0.43\pm 6.09$   \\
\hline
   $N(1720)\rightarrow \pi \ N$
    & $22.5 \pm 11$   &   28.5    &   28.4   \\
   $N(1720)\rightarrow \pi \ \Delta$
   &  unknown         &    2.2    &   2.4   \\
   $N(1680)\rightarrow \pi \ \Delta$
   & $13 \pm 5$       &    11.9   &  11.5    \\
   $\Delta(1910)\rightarrow \pi \ N$
   & $52\pm 20$       &    46.2  &   48.1   \\
   $\Delta(1910)\rightarrow \pi \ \Delta$
   &  unknown         &    5.25   &    5.9    \\
   $\Delta(1920)\rightarrow \pi \ N$
   & $28\pm 19$       &    25.1  &    24.7   \\
   $\Delta(1920)\rightarrow \pi \ \Delta$
   & unknown          &    19.3  &   20.3    \\
   $\Delta(1905)\rightarrow \pi \ \Delta$
   & unknown          &    22.0   &   21.1    \\
\hline
\hline
\end{tabular}
\end{center}
\end{table}

\begin{table}[h]
\begin{center}
\caption{Fit parameters and partial widths corresponding to the pion F wave decays of
$\ell=2$ excited baryons. Note that for the LO fit those empirical errors that are
less than 30\% have been increased up to that value. In the fit \#2 NLO the empirical
value of the decay $N(1680)\rightarrow \pi \ \Delta$ has not been considered.}
\vspace*{1cm}\begin{tabular}{ccccc}
\hline\hline
Pion F waves  & \hspace*{.5cm} Emp. Width   \hspace*{.5cm} &  \hspace*{.5cm} LO   \hspace*{.5cm}
& \hspace*{.5cm} \#1 NLO   \hspace*{.5cm} & \hspace*{.5cm} \#2 NLO \hspace*{.5cm} \\
              &     MeV        &   MeV         &   MeV    &   MeV    \\
\hline
$\chi^2_{dof}$&                &     1.73      &   6.9    &  0.38     \\
dof           &                &      4        &     2    &  1       \\
\hline
$C_1^{[3,1]}$ &                & $1.01\pm 0.09$  & $0.78\pm 0.04$  &  $0.84\pm 0.04$  \\
$C_2^{[3,1]}$ &                &      -          & $-0.84\pm 0.18$ &  $-1.30\pm 0.21$ \\
$C_3^{[3,1]}$ &                &      -          & $-0.81\pm 0.26$ &  $-0.26\pm 0.47$  \\
\hline
   $N(1720)\rightarrow \pi \ \Delta$
    &     unknown     &    6.86     &    19.9    &   25.1  \\
   $N(1680)\rightarrow \pi \ N$
   &  $84.5\pm 9$     &    44.8     &    72.9    &   86.0  \\
   $N(1680)\rightarrow \pi \ \Delta$
   &    $1.3\pm1.3$   &    2.39     &     4.84   &    7.94 \\
   $\Delta(1920)\rightarrow \pi \ \Delta$
   &     unknown      &    17.3     &     51.4   &    40.4 \\
  $\Delta(1905)\rightarrow \pi \ N$
   & $36\pm 20$       &   26.0     &     43.5   &    25.9 \\
  $\Delta(1905)\rightarrow \pi \ \Delta$
   &     unknown      &   24.8      &    51.6   &    51.3 \\
  $\Delta(1950)\rightarrow \pi \ N$
   &  $120\pm 14$     &  159      &    129   &   121 \\
  $\Delta(1950)\rightarrow \pi \ \Delta$
  &   $77\pm15$       &  42.0      &     46.3   &    72.3 \\
\hline
\hline
\end{tabular}
\end{center}
\end{table}

\end{document}